\documentclass[amsmath,amssymb,superscriptaddress]{revtex4-1}
\usepackage{graphicx}
\usepackage{dcolumn}
\usepackage[mathlines]{lineno}
\usepackage{hyperref}
\usepackage{bm}
\usepackage{epstopdf}
\usepackage{epsfig}
\usepackage{amsmath}
\usepackage{units}

\usepackage{color}

\newcommand{\fat}{\boldsymbol}

\begin{document}


\title{Angular acceleration with twisted light}

\author{Christian Schulze}
\affiliation{Institute of Applied Optics, Friedrich Schiller University, Fr\"{o}belstieg 1, 07743 Jena, Germany} 
\author{Filippus S.~Roux} 
\affiliation{National Laser Centre, Council for Scientific and Industrial Research, P.O. Box 395, Pretoria 0001, South Africa}
\author{Angela Dudley} 
\affiliation{National Laser Centre, Council for Scientific and Industrial Research, P.O. Box 395, Pretoria 0001, South Africa}
\author{Ronald Rop}
\affiliation{Department of Physics, Egerton University, P.O. Box 536, Egerton 20115, Kenya.}
\author{Michael Duparr\'{e}}
\affiliation{Institute of Applied Optics, Friedrich Schiller University, Fr\"{o}belstieg 1, 07743 Jena, Germany} 
\author{Andrew Forbes}
\affiliation{National Laser Centre, Council for Scientific and Industrial Research, P.O. Box 395, Pretoria 0001, South Africa}
\affiliation{School of Physics, University of the Witwatersrand, Private Bag X3, Johannesburg 2030, South Africa}
\affiliation{Corresponding email: aforbes1@csir.co.za}
\date{\today}

\begin{abstract}
\textbf{\noindent There is significant interest in tailoring wave packets that transversely accelerate during propagation\cite{bandres2013}. The first realisation was in the optical domain, the Airy beam\cite{siviloglou2007}. Valid in the paraxial approximation, such beams were shown to transversely accelerate while following a parabolic caustic, even if the centroid itself travelled along a rectilinear path. Later, non-paraxial optical fields in the form of Weber and Mathieu beams were demonstrated\cite{zhang2012}, as well as non-optical wave packets with electrons\cite{voloch2013}.  Such fields have found a plethora of applications from particle manipulation\cite{baumgartl2008}, spatial-temporal beam control\cite{chong2010}, plasma control\cite{polynkin2009} to non-linear optics\cite{ellenbogen2009}. Here we demonstrate, for the first time, the angular acceleration of light, achieved by non-canonical superpositions of Bessel beams carrying orbital angular momentum. We show that by adjustment of a single parameter the acceleration and deceleration may be tuned continuously. We observe an unexpected energy transfer mechanism between spatial regions of the field which we investigate experimentally and theoretically.  Our findings offer a new class of optical field that will enable studies in matter waves and opto-fluidics.}       
\end{abstract}

\maketitle

\noindent The concept of accelerating light at first appears incongruous with the fact that light travels at a constant speed and in a straight line. But it has been shown that if specific features of the fields are considered in isolation, then strange and counterintuitive propagation characteristics can be realised.  The most recent examples all consider features of the field that appear to transversely accelerate as they propagate.  These include the now well-known Airy beams\cite{siviloglou2007}, whose intensity peak follows a parabolic path through space even if the centroid itself obeys rectilinear propagation.  Such fields suffer from rapid deviation from the paraxial approximation due to the constantly changing propagation angle.  Thus while they exhibit interesting transverse acceleration, which has seen them applied in a range of fields, this deviation from the paraxial approximation is a limiting factor.  Consequently they do not hold their shape for very long, have limited transverse acceleration, are restricted in feature sizes that can be realised and do not lend themselves to applications requiring a sharp focus (or large angles). More recently two-dimensional parabolic accelerating beams have been found to overcome some of these limitations, as well as nonparaxial transversely accelerating beams in the form of Weber\cite{bandresnjp2013,zhang2012}, Mathieu\cite{zhang2012} and Bessel beams\cite{chremmos2013}, as well as vector\cite{aleahmad2012} and arbitrary shaped\cite{ruelas2014} manifestations of the same.

Here we outline and demonstrate \textit{angular} accelerating light.  We create superpositions of non-diffracting, high-order Bessel beams\cite{McGloin2012,mazilu} carrying orbital angular momentum (OAM) \cite{Yao2011}, and tailor the phase to have a tuneable non-linear variation with azimuthal angle.  We show that the degree of non-linearity determines the magnitude of the acceleration and deceleration, and that it may readily be tuned by adjusting a single parameter.  Conveniently, since the acceleration is not directly coupled to the feature sizes within the field, there is no limit on the amount of angular rotation that can be tolerated, and consequently the angular acceleration/deceleration may continue for arbitrarily long distances. We create such fields using computer generated holograms and discover a new energy exchange mechanism not previously observed, which we explore both experimentally and theoretically.

\begin{figure}[h]
\includegraphics{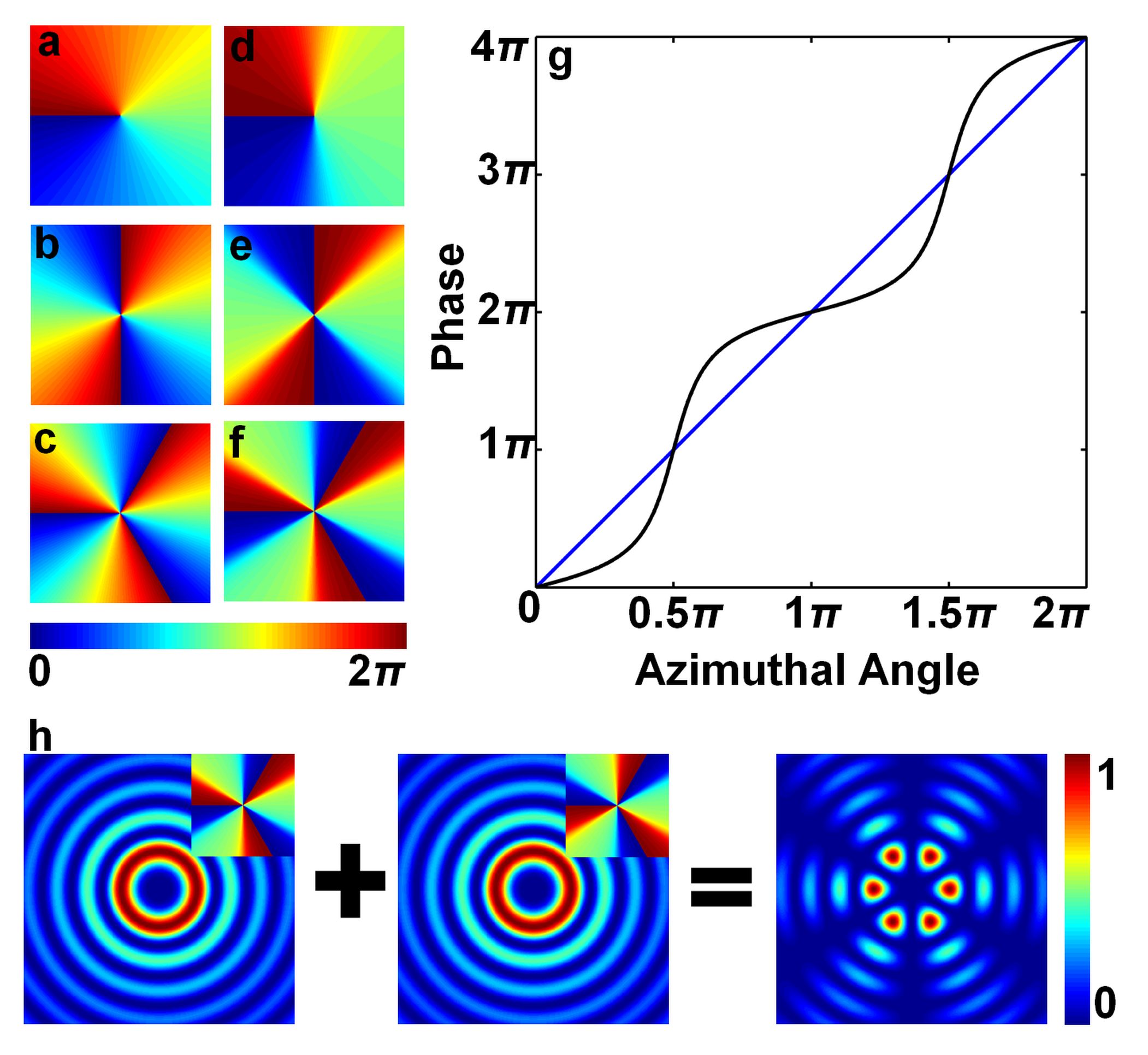}
\caption{\textbf{Canonical and non-canonical vortices.} (a)-(c) Canonical vortex phases for $\ell = 1,2$ and $3$, respectively. (d)-(f) Non-canonical vortex phases for $\ell = 1,2$ and $3$, respectively, (g)  A plot of the phase as a function of azimuthal angle for a canonical (blue) and non-canonical (black) vortex of order $\ell = 2$. (h) Schematic showing the superposition of two non-canonical Bessel beams as intensities (with vortex phase as insets) and the resulting intensity petal structure, for $\ell = 3$.  All non-canonical plots are for $\theta = \pi/3$.}
\label{canonical}
\end{figure}

To begin we recall that light fields with a phase term of $\exp(i\ell \varphi)$, where $\varphi$ is the azimuthal angle and $\ell$ the topological charge of the field, may carry orbital angular momentum of $\ell \hbar$ per photon\cite{Yao2011}.  Such fields are variously referred to as vortex or twisted light because of the helical wavefronts of helicity $\ell$. Examples of such fields are found as the familiar Laguerre-Gaussian and Bessel-Gaussian beams, and have been studied and applied in both the classical and quantum regimes\cite{Yao2011}.  We notice that there is a phase singularity (and intensity null) at the centre of the beam where a vortex of order $\ell$ is to be found.  \textbf{Figures \ref{canonical} (a)-(c)} show examples of such vortex beams for $\ell=1,2$ and $3$. These are canonical vortex fields, and pertinently, the phase varies linearly with the azimuthal angle, with slope $\ell$. We will show that a key requirement for angular acceleration is a field where the phase varies in a non-linear fashion about the azimuth. It is not possible to engineer this phase variation arbitrarily (e.g., as some power law of the azimuthal angle) since the periodic nature of the phase in OAM fields demands that the non-linearity is also periodic.  However non-canonical vortex fields ensure the periodicity, with some examples shown in \textbf{Figures \ref{canonical} (d)-(f)}. Plotting both the linear (canonical) and non-linear (non-canonical) phase variations together [\textbf{Figures \ref{canonical} (g)}] we note that both are periodic about the azimuthal plane.  To create such fields we point out that a non-canonical vortex can always be expressed as the linear combination of opposite helicity canonical vortex beams, which at $z=0$ may be written as

\begin{equation}
u_{\rm nc}(r,\varphi,\theta)  =  A_{\ell}(r) [\cos(\theta/2)\exp(i \ell \varphi) + \sin(\theta/2)\exp(-i \ell \varphi)],
\label{ncu} 
\end{equation}

\noindent where $A_{\ell}(r)$ is some radial ($r$) enveloping function and $\theta$ determines the morphology (anisotropy) of the optical vortex on the axis of the beam. The morphology parameter $\theta$ governs the relative weights of the two opposite topological charges such that the overall intensity remains constant irrespective of the value of $\theta$. For $0<\theta<\pi$ the overall topological charge is positive and for $\pi<\theta<2\pi$ it is negative.  By adding two such non-canonical vortex beams of differing phase velocities ($k_z$) and opposite morphology, the resulting field 

\begin{equation}
u(r,\varphi,z)  =   u_{\rm nc}(r,\varphi,\theta) \exp(i k_{z1} z) + u_{\rm nc}(r,\varphi,\pi - \theta) \exp(i k_{z2} z),
\label{ncu} 
\end{equation}

\noindent will have a structured pattern in the azimuth ($\varphi$) that rotates during propagation, with an angular velocity that is dependent on $z$ (see also Supplementary Information). This is shown schematically in \textbf{Figure \ref{canonical} (h)} where the two non-canonical fields result in a petal-like structure about the azimuthal plane.  

\begin{figure}[h]
\includegraphics{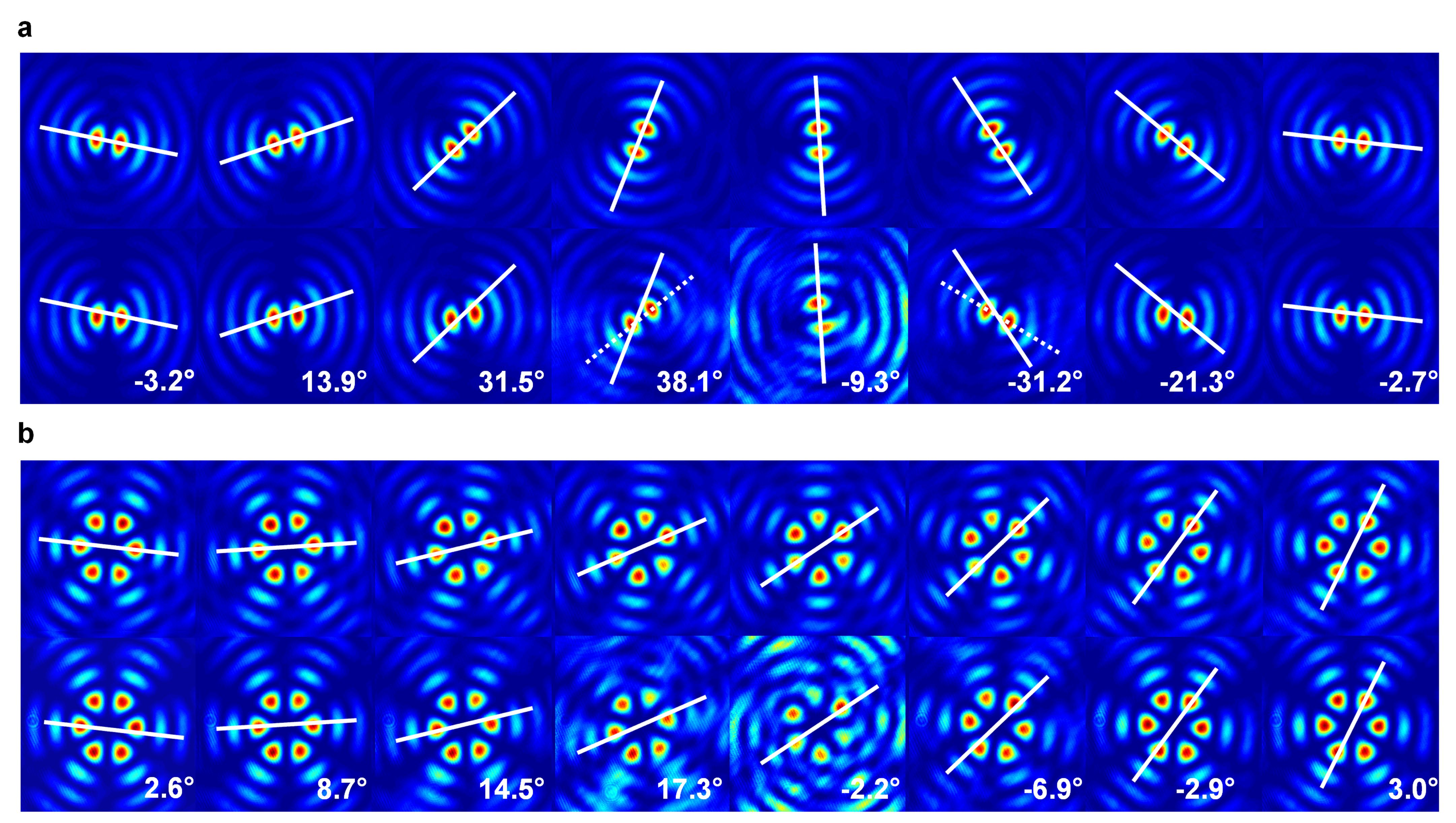}
\caption{\textbf{Comparison of linear and non-linear rotation.} The intensity images in the top rows of (a) and (b) show the petal rotation at a fixed rate, $\theta = 0$, while the bottom rows show the non-linear rotation, $\theta = \pi /3$, for (a) $\ell = 1$ and (b) $\ell = 3$. The solid white lines indicate the movement of the petals in the linear case and are overlaid on the non-linear images for reference.  The angular difference between the two cases (linear and non-linear) is given as text in each frame. The columns represent increasing propagation distances sampled from the zoomed in data set of \textbf{Figure \ref{angrot}}. The maximum beam intensities in each frame were normalised to unity to aid comparison.}
\label{rotframes}
\end{figure}

For convenience, and without any loss of generality, we will discuss the concept and implementation thereof in the context of Bessel beams as the spatial mode so that $A_{\ell}(r) ={\rm J}_{\ell}(r k_{r})$.  Such fields are convenient as their radial wavevector ($k_r$) and phase velocity ($k_z$) are easily controlled with digital holograms (see Supplementary Information), and are related by 
\begin{equation}
k_z = \sqrt{k^2-k_r^2},
\label{kzkr} 
\end{equation}

\noindent where $k=2\pi / \lambda$ is the wavenumber and $\lambda$ is the wavelength of the light.  Our superposition field then becomes

\begin{eqnarray}
u(r,\varphi,z)  & = & {\rm J}_{\ell}(r k_{r1}) [\cos(\theta/2)\exp(i \ell \varphi) + \sin(\theta/2)\exp(-i \ell \varphi)] \exp(i k_{z1} z) \nonumber \\ & & + {\rm J}_{\ell}(r k_{r2}) [\sin(\theta/2)\exp(i \ell \varphi) + \cos(\theta/2)\exp(-i \ell \varphi)] \exp(i k_{z2} z),
\label{supu}
\end{eqnarray}
        
\begin{figure}[h]
\includegraphics{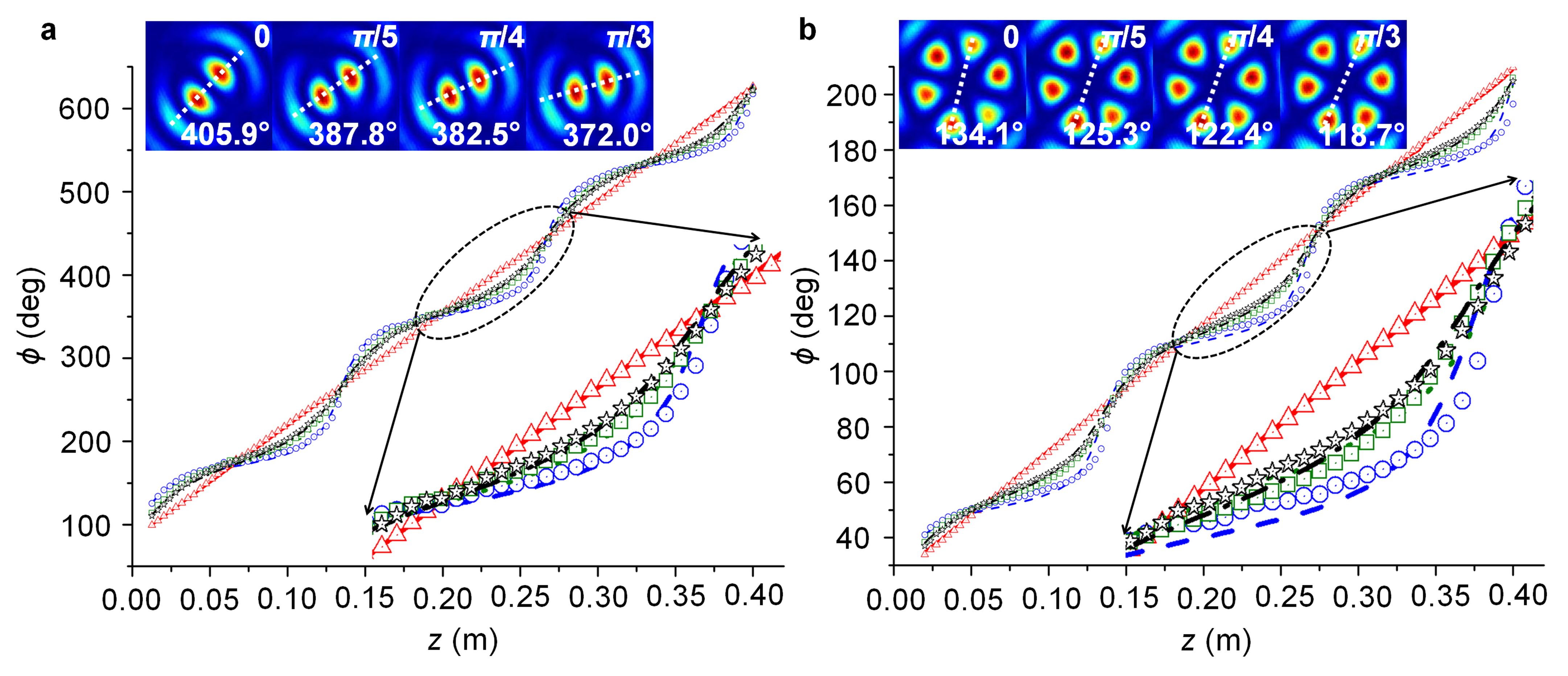}
\caption{\textbf{Angular rotation of the petals during propagation.} The angular rotation ($\phi$) of the petal structure in degrees as the field propagated over a distance ($z$) of approximately half a metre.  The measured data (symbols) is shown together with the theoretical predictions (dashed lines) for various values of the control parameter, $\theta$, from 0 (red), $\pi/5$ (black), $\pi/4$ (green) to $\pi/3$ (blue).  The rotation rate was observed to transition from constant ($\theta = 0$) to highly non-linear ($\theta = \pi /3$).  Results are shown for two different topological charge values of (a) $\ell = 1$ and (b) $\ell = 3$.  The zoomed in sections show that the experimental data fits the theoretical predictions very well.  The image insets show the beam profile at a distance of $z = 236.6$ mm (a position at which maximum angular deviation from the linear case is observed) for the four plotted $\theta$ values, with the angular rotation of the petals given as white text.}
\label{angrot}
\end{figure}

The optical field in equation (\ref{supu}) consists of the superposition of four Bessel beams --- two pairs with slightly different values of $k_r$ (and therefore slightly different values of $k_z$). Each pair is a superposition of two Bessel beams with opposite azimuthal index that produces an anisotropic (non-canonical) optical vortex in the centre of the beam.  This ensures both requirements: non-linear but periodic phase variation around the azimuthal plane, as shown in \textbf{Figure \ref{canonical} (g)}.  The spatial structure of this accelerating field consists of petal-like intensity lobes about the azimuth, \textbf{Figure \ref{canonical} (h)}, where the number of petals is given by $2|\ell|$.  If we consider the rotation of any particular petal, it is easy to show (see Supplementary Information) that the angular position ($\phi$) rotates during propagation following    

\begin{equation}
\phi(z) = \frac{1}{2|\ell|} \arctan \left[ \frac{\cos(\theta)\sin(z\Delta)}{\sin(\theta) + \cos(z\Delta)} \right],
\label{angle}
\end{equation}

\noindent where $\Delta = k_{z2} - k_{z1}$.  This rotation was observed experimentally using a visible laser source and a spatial light modulator (see Methods) encoded with digital holograms to produce the desired field given by equation (\ref{supu}).  The intensity patterns were recorded during propagation and are shown for the linear and non-linear cases in \textbf{Figures \ref{rotframes} (a)} and \textbf{(b)} for superpositons of $\ell = 1$ and $\ell = 3$, respectively (see Supplementary Movies).  The predicted and measured angular position during propagation for selected values of $\theta$ are shown in \textbf{Figures \ref{angrot} (a)} and \textbf{(b)}, for $\ell = 1$ and $\ell = 3$, respectively. The theoretical predictions are validated by the experimental results.   

We note that the morphology parameter, $\theta$, acts as a tuning parameter that determines the degree of non-linearity, and therefore also the angular velocity of the rotation.  The latter can be shown to be 

\begin{equation}
\frac{\partial \phi}{\partial z} = \dot{\phi} = \frac{\Delta}{2|\ell|} \frac{\cos(2\theta)}{1 + \sin(2\theta) \cos(z\Delta)},
\label{rotrate}
\end{equation}

\noindent from which we may immediately find the angular acceleration:

\begin{equation}
\frac{\partial \dot{\phi}}{\partial z} = \ddot{\phi} = \frac{\Delta^2}{4|\ell|} \frac{\sin(2\theta) \sin(z\Delta)}{[1 + \sin(\theta) \cos(z\Delta)]^2}.
\label{accrate}
\end{equation}

\begin{figure}[ht]
\includegraphics{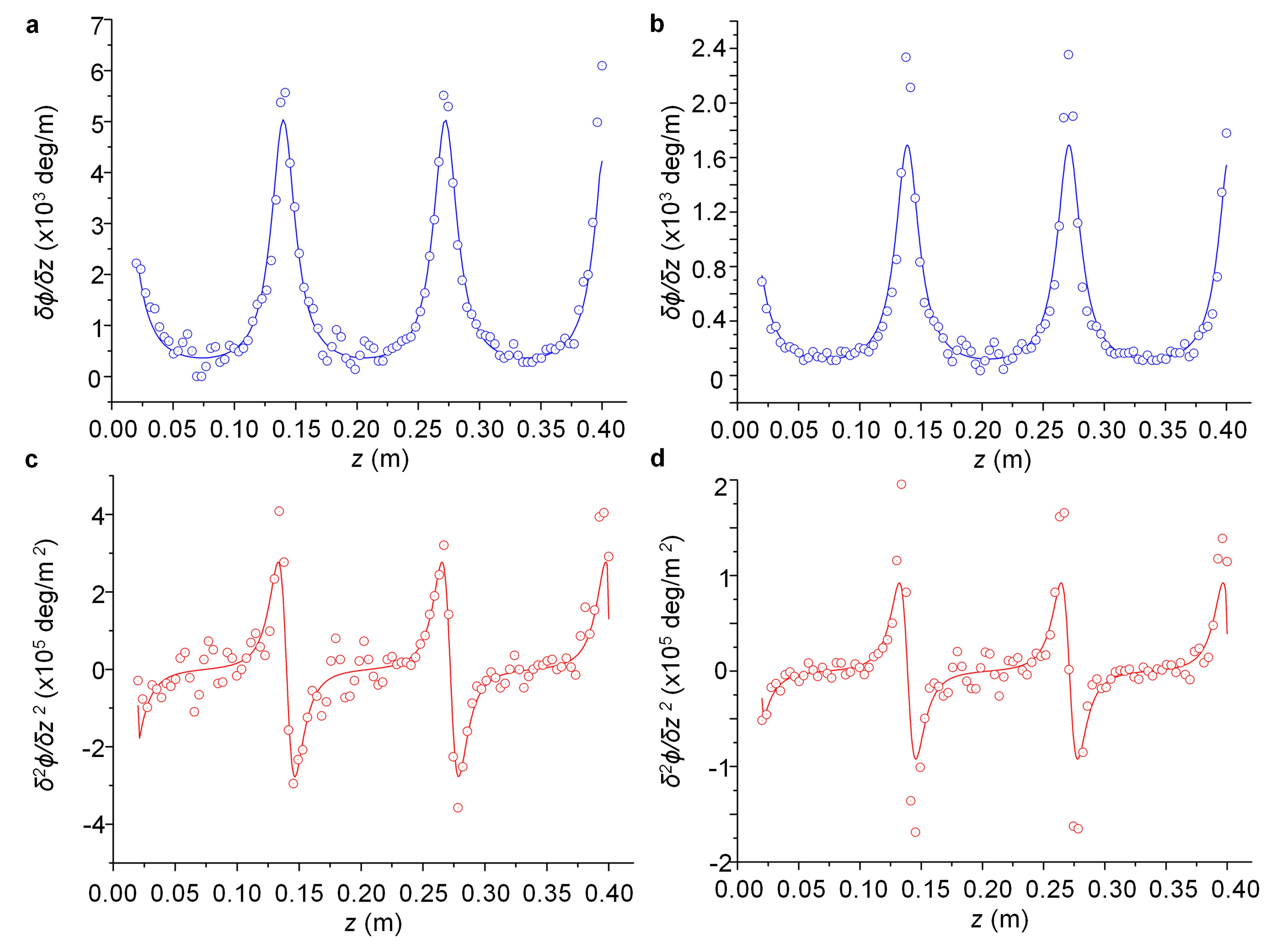}
\caption{\textbf{Angular velocity and acceleration of the rotating fields.} The non-linear azimuthal phase of the non-canonical beam results in a rotation angle that varies non-linearly with propagation distance.  This varying angular velocity imparts angular acceleration to the field.  The non-constant angular velocity is shown in (a) and (b) for $\ell = 1$ and $\ell = 3$, respectively, while the resultant angular acceleration for these respective examples is shown in (c) and (d).  The measured data (circles) is in good agreement with the theoretical predictions (solid curve).}
\label{accelerate}
\end{figure}

These new relations have been confirmed experimentally, with results shown in \textbf{Figure \ref{accelerate}} for $\ell = 1$ and $\ell = 3$.  As the field propagates in the positive $z$ direction, so the petal pattern rotates.  The angular velocity of this rotation changes during propagation according to equation (\ref{rotrate}), following an oscillatory evolution.  This is evident in \textbf{Figures \ref{accelerate} (a)} and \textbf{(b)} for $\ell = 1$ and $\ell = 3$ beams, respectively.  The velocity changes rapidly as the beam propagates, speeding up and slowing down, reaching very high angular velocities.  

This changing angular velocity implies an angular acceleration, the origin of which resides in the phase structure of the superposition field.  As the angular velocity is oscillatory, so the beam experiences both acceleration and deceleration during propagation.  A plot of this can be seen in \textbf{Figures \ref{accelerate} (c)} and \textbf{(d)} for the $\ell = 1$ and $\ell = 3$ helicities.  The magnitude of the angular velocity and angular acceleration is affected by both the helicity of the beams in the superposition ($\ell$) as well as the difference in their phase velocities ($\Delta$), following equations (\ref{rotrate}) and (\ref{accrate}).  The tuning (morphology) parameter, $\theta$, determines if the field will accelerate at all, and to what extent. Previous studies of rotating fields can be deduced by setting $\theta = 0$; the petals rotate at a constant angular velocity given by $\phi=z\Delta/2\ell$ since equation~(\ref{ncu}) reduces to a superposition of two OAM fields with linear azimuthal phase variation.  Such fields, \textit{sans} acceleration, have been studied in detail previously\cite{rot1,rot2,rot3,rot4,rot5,rot6,rot7,rop1} and represent only a special case of the more general accelerating light described here.

\begin{figure}[h]
\includegraphics{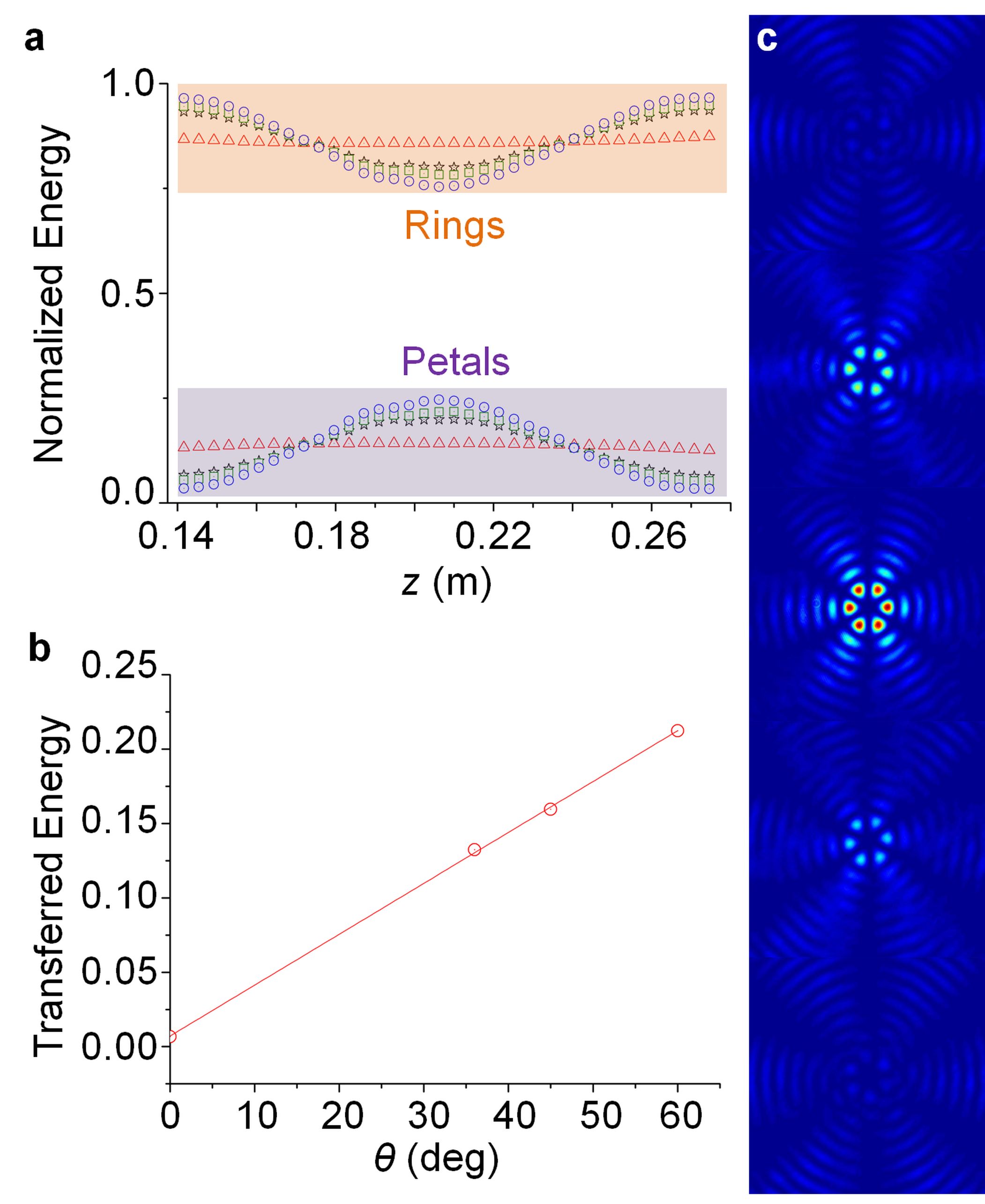}
\caption{\textbf{Energy transport during acceleration and deceleration.} As the beam accelerates so the energy in the inner region (petals) is transported to the outer region (rings).  This process is reversed during deceleration.  Because the field oscillates between the two cases during propagation, there is a continuous flow of energy in and out of the central petal structure of the field. (a) The normalised energy in the two regions (petals and rings) of the field during propagation for $\theta = 0$ (red), $\theta = \pi/5$ (black), $\theta = \pi/4$ (green) and $\theta = \pi/3$ (blue). (b) The measured maximum fractional energy exchange as a function of $\theta$, and (c) beam snapshots (unnormalised) over a complete cycle at distances of approximately $z= 140, 180, 210, 240$ and $270$ mm.  All data is for the $\ell = 3$ case. }
\label{energy}
\end{figure}

Up to this point our analysis has been restricted to the rotation of the petal structure.  A new and interesting feature of these angular accelerating beams becomes evident during propagation when the entire field is considered: the central petal structure appears to gain and lose energy in a manner that is directly coupled to the acceleration.  When the inner petal structure is accelerating, energy is transferred out of this region of the field, whereas when it is decelerating energy is transferred back into the petal structure. In other words, a fast rotating field appears dim, while a slow rotating field appears bright.  This exchange is measured and plotted in \textbf{Figures \ref{energy} (a)} and \textbf{(b)}, with the raw beam data over one cycle shown in \textbf{Figure \ref{energy} (c)}. We have analysed this problem theoretically (see Supplementary Information) and find that the energy is transported radially during propagation due to the fact that the entire field is not rotating at the same rate.  Indeed, while our prior analysis pertains to the central region of the field (petals), the outer region is also rotating but exactly out of phase with the inner region.  We find that the angular accelerations are inversely proportional to their $z$-dependent intensity modulations, so that the two regions exchange energy and alternate in brightness during propagation. A fast rotation in one part of the field is accompanied by a slow rotation in the other part.  In summary, each of the two parts of the intensity pattern rotates with a varying rotation rate. When one rotates faster, the other rotates slower. Since the rotation rates are inversely proportional to their respective overall intensities, the one that rotates slow is always brighter than the other one, which rotates fast (see Supplementary Movies). 

Finally, we point out that we have implemented the concept with Bessel beams for convenience only: Bessel beams may have their phase velocities engineered by the radius of a ring-slit programmed on a spatial light modulator.  In principle \textit{any} field that satisfies the requirements outlined here will result in angular acceleration.  Our treatment may be generalised further to create Helicon beams \cite{rop2} that accelerate by not fixing the superpositions to equal but opposite helicities.  While many studies have considered rotating light created by a myriad of techniques, none have demonstrated controlled angular acceleration.  Such fields will have obvious applications in the optical control of micro-particles \cite{Gluckstad2012,Kishan2011,Miles2011} and may even be extended to the non-optical domain \cite{Karimi2014} and non-linear propagation \cite{Kivshar2005, Soskin2006} to explore new physical processes.

In conclusion, we have outlined a new concept for the angular acceleration of light and demonstrated it experimentally.  Our approach makes use of superpositions of OAM fields created with digital holograms. We have shown that these angular accelerating fields may be tuned continuously in their acceleration with the aid of a single morphology parameter.  These fields have advantageous features that overcome previous disadvantages of transversely accelerating light: their feature sizes are not dictated by the degree of acceleration, the angular extent of the acceleration does not influence the paraxial nature of the field (the entire azimuthal plane can be used) and they can be engineered to extend over arbitrarily long distances.  Given the interest and applications of transversely accelerating light, one can envisage many uses for this new class of angular accelerating light, for example, in driving flow of opto-fluidic systems and accelerating matter waves studies. 

\section*{Methods}
\textbf{Experimental details.} The laser source was a linearly-polarized, single wavelength ($\lambda \sim$ 633 nm) helium-neon laser (Melles Griot) with a power of $\sim$ 10 mW which was expanded and collimated by a telescope ($f_{\textrm{L1}}$ = 15 mm and $f_{\textrm{L2}}$ = 125 mm) to approximate a plane wave. The plane wave illuminated a HoloEye Pluto spatial light modulator (SLM) ($1080 \times 1920$ pixels) which has a resolution of 8 $\mu$m and was calibrated for a wavelength of 633 nm. The SLM was addressed with holograms representing Durnin's ring-slit aperture encoded via amplitude modulation (see Supplementary Information). Two ring-slit apertures with $R1 = 179$ pixels (1432 $\mu$m), $R2 = 195$ pixels (1560 $\mu$m) and $d = 16$ pixels (128 $\mu$m) were used, each encoded with a non-canonical azimuthal phase variation of order $\ell$ and $-\ell$, respectively, so that the transmission function on the SLM was given by

\begin{equation}
t(r, \varphi) = \left\{ 
\begin{array}{l l}
\cos(\theta/2)\exp(i \ell \varphi) + \sin(\theta/2)\exp(-i \ell \varphi) & \quad R1-d/2 \leq r \leq R1 +d/2\\
\sin(\theta/2)\exp(i \ell \varphi) + \cos(\theta/2)\exp(-i \ell \varphi) & \quad R2-d/2 \leq r \leq R2 +d/2\\
0 & \quad \text{elsewhere}
\end{array} \right. .
\end{equation}

The Fourier transform of the field at the plane of the SLM (i.e. the superposition of two Bessel beams of orders $\ell$ and $-\ell$ with $k_{r1} = 28428.1$ $\textrm{m}^{-1}$  and $k_{r2} = 30969.2$  $\textrm{m}^{-1}$) was achieved with the use of a lens: $f_{\textrm{L3}}$ = 500 mm and magnified with a $10 \times$ objective. The $10 \times$ objective also acted as an aperture to select only the first diffraction order which was captured on a CCD camera (Point Grey fire-wire CCD).

\textbf{Petal position.} To extract the petal position from the captured camera images, the measured Bessel beams were compared to simulated beams with the same number of petals, which were adapted in spatial scale. The comparison was achieved in a quantitative manner by evaluating a two-dimensional cross-correlation coefficient when rotating the simulated patterns with respect to the measured beam. The correct angular rotation of the measured beam was found from a maximum of the correlation function. To avoid ambiguities in the form of a multiple number of correlation maxima, the simulated pattern was rotated within the interval of 0$^\circ$ to 180$^\circ / \ell$.  The fidelity of the process was improved by iteratively refining the interval limits until a chosen accuracy was achieved, which was 0.1$^\circ$ in the experiments.

\textbf{Power exchange.} For each beam intensity $I$ recorded at a propagation distance $z$, the center, which was obtained from the first order moments, was surrounded by a circle, whose radius $R$ was manually adapted to enclose the petal region. From integrating the beam intensity inside and outside of this circular aperture, the power of the petal and ring region was determined, and normalized to the respective total power $P_\text{tot}$:
\begin{eqnarray}
P_\text{petal}&=&\frac{1}{P_\text{tot}}\iint_{x^2+y^2<R^2}I(z)dxdy\\
P_\text{rings}&=&\frac{1}{P_\text{tot}}\iint_{x^2+y^2>R^2}I(z)dxdy,
\end{eqnarray}
which is necessary due to the limited definition region of the Bessel beams (see Supplementary Information). The power exchange between the petal and the ring region is then identified from the difference of the maximum to the minimum of the petal power $P_\text{petal}^\text{max}-P_\text{petal}^\text{min}$.

\section*{Acknowledgements}
A.F. would like to thank the National Research Foundation of South Africa for financial support under grant 78977. 
\section*{Author contributions}
C.S., A.D. and R.R. performed the experiments at the CSIR National Laser Centre, F.S.R contributed the theoretical analysis and all authors contributed to the analysis of the results.  A.F. conceived of the idea and wrote the manuscript, with inputs from all authors.
\section*{Additional information}
See Supplementary Information and Movies.
\section*{Competing financial interests}
The authors declare no competing financial interests.

\newpage
\appendix
\section*{Supplementary Information}
\label{SI}

\subsection{Encoding the non-canonical Bessel beams by complex amplitude modulation}
Bessel beams are created from a superposition of plane waves with wave vectors that lie on a cone and are infinite energy, non-diffracting solutions to the Helmholtz equation in cylindrical coordinates\cite{Durnin1987,Durnin21987}.  Such fields are approximated to a very good degree by Bessel-Gaussian beams\cite{Gori1987} given by 
\begin{equation}
E^{\rm BG}_{\ell}(r,\varphi,z)  =  \sqrt{\frac{2}{\pi}} {\rm J}_{\ell} \left(\frac{ z_{\textrm{R}} k_r r}{z_{\textrm{R}}-iz}\right) \exp(i\ell\varphi-ik_z z) \exp \left( \frac{i k_r^2 z w_0^2 - 2 k r^2}{4 (z_{\textrm{R}}-iz)} \right),
\label{bgm}
\end{equation}
where $\ell$ is the azimuthal index, ${\rm J}_{\ell}(\cdot)$ is the Bessel function of the first kind, and $k_{r}$ and $k_{z}$ are the radial and longitudinal wave vectors, respectively. The initial radius of the Gaussian profile is $w_0$ and the Rayleigh range is $z_{\textrm{R}}=\pi w_0^2/\lambda$, where $\lambda$ is the wavelength of the illuminating light.  

\begin{figure}[h]
\includegraphics{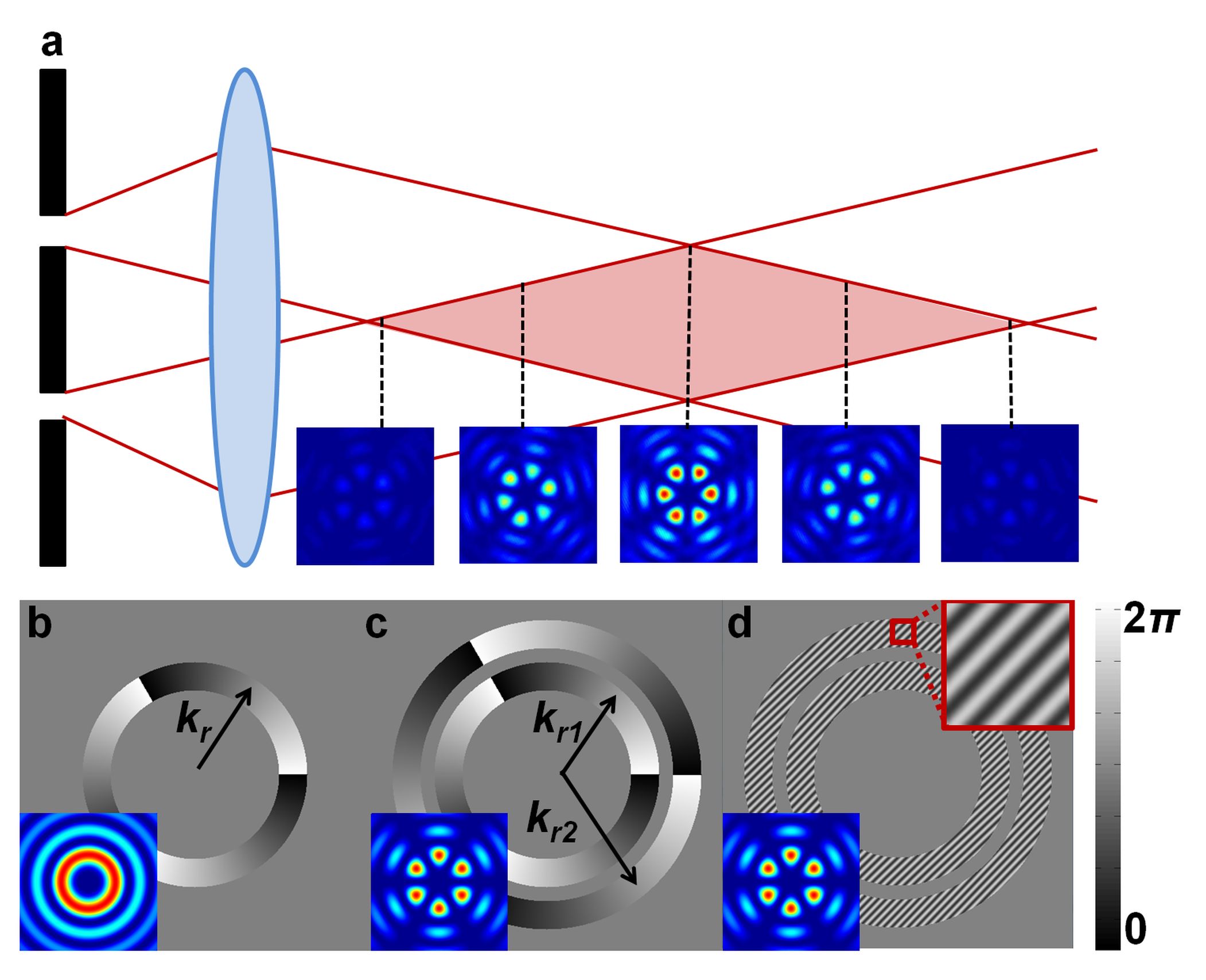}
\caption{\textbf{Bessel beam generation.} (a) A Bessel beam may be formed by illuminating a ring-slit placed in the Fourier plane of a lens. (b) To create a field comprising a single radial wavevector, $k_r$, a single ring-slit is illuminated. Superpositions of canonical Bessel beams with differing radial wavevectors simply requires multiple ring-slits as shown in (c), while non-canonical superpositions require amplitude modulation inside each of these ring slits, as shown in (d).  The corresponding intensities produced for each case are shown in the insets.}
\label{concept}
\end{figure} 

We create the Bessel beams using a modified version of Durnin's original ring-slit experiment, where the Bessel beam approximation is valid over a finite propagation distance, $z_{\textrm{max}}=2\pi\omega_{0}/\lambda k_{r}$, shown as the shaded red region in \textbf{Figure \ref{concept} (a)}. In this arrangement the ring-slit is programmed as a computer generated hologram encoded on a spatial light modulator (SLM) and placed in the Fourier plane of a lens\cite{Turunen1988,dudley}.  As such, the radius of the ring on the hologram determines $k_r$ and consequently $k_z$.  Introducing a hologram with two ring-slits with different radial wave-vectors producing a superposition of Bessel beams with differing phase velocities.  The azimuthal phase is varied inside the ring to impart OAM to the field, and to fix the order of the Bessel beams.  

Examples of these holograms and the resulting fields are shown in \textbf{Figure \ref{concept} (b) and (c)}.  To create the non-canonical superposition requires control of both amplitude and phase on the SLM. This can be achieved by encoding the ring-slit holograms via type III complex amplitude modulation\cite{arrizon}, where  a phase-only function, $\exp[i\psi(r,\varphi)]$, is determined from the ansatz $\psi(r, \varphi) = f(|U|)\sin[\arg(U)]$, and $J_1[f(|U|)] = 0.58|U|$, where $U$ is the near-field of the beam of interest (i.e. the ring-slit).  An example of such a hologram is shown in \textbf{Figure \ref{concept} (d)}. Superimposing a grating allows the beam of interest to be separated from the  zero order with the use of an aperture placed in the far-field plane of the hologram. 


\subsection{Virtual propagation of the fields}
\begin{figure}[h]
\includegraphics[width=10cm]{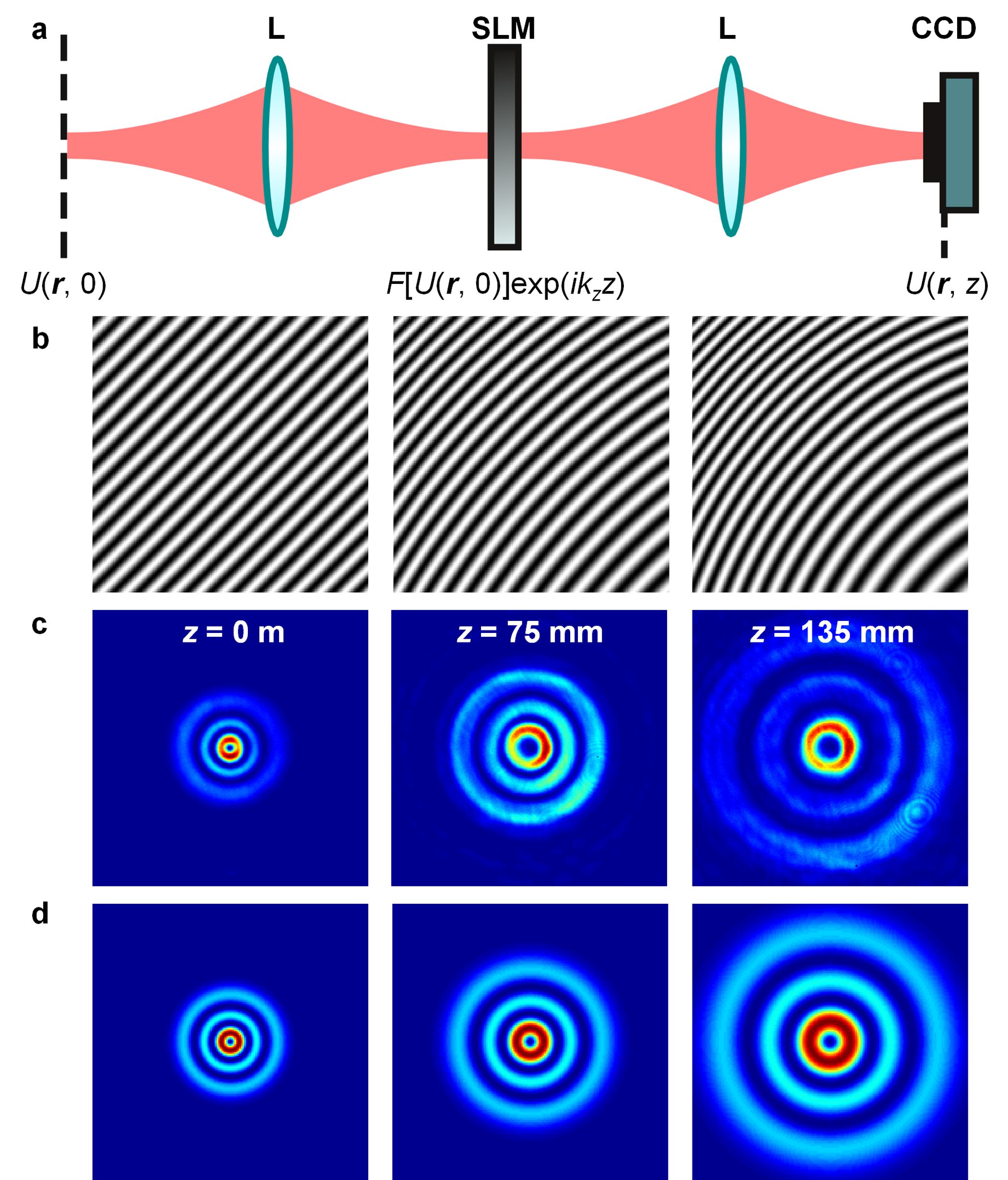}
\caption{\textbf{Digital propagation concept.}(a) Scheme of the digital propagation using lenses $L$, a spatial light modulator (SLM), and a camera (CCD), which records the digitally propagated beam. (b) Digital holograms displayed on the SLM (lens focal length $f=\unit[200]{mm}$). The grating period was enlarged for better illustration. c) Laguerre Gaussian beam ($p=2,\ell=1$,) with a fundamental mode waist diameter 0.4 mm, after diffraction at the holograms yielding digital propagation over distances of  0 mm, 75 mm, and 135 mm. (d) The simulated propagation of the beam for the same distances. Intensities depicted normalized to the respective maximum values.}
\label{S2}
\end{figure}

We propagate our field using a motion-free approach that employs a spatial light modulator to mimic the angular spectrum method for beam propagation \cite{christian}. According to the angular spectrum approach an optical field can be regarded as a superposition of plane waves, which are spatial Fourier components, and travel in different directions. The propagation of such a field $U(\fat{r},z=0)$ in the direction of the $z$-axis can be described with the help of the transfer function of free space $\exp(ik_z z)$:
\begin{equation}
U(\fat{r},z)=\mathcal{F}^{-1}\{\mathcal{F}[U(\fat{r},0)]\exp(ik_z z)\},
\label{eq:dp}
\end{equation}
where $\mathcal{F}$ and $\mathcal{F}^{-1}$ denote the Fourier transform and its inverse, respectively, and $k_z$ is the $z$-component of the wave vector. In a physical picture, the Fourier transforms can be realized by lenses, where the focal length $f$ determines the wave vector components: $k_z=2\pi/\lambda(1-x^2/f^2-y^2/f^2)^{1/2}$. The phase function $\exp(ik_z z)$ can be implemented by digital means, e.g., with an SLM. The experimental realization of equation~\eqref{eq:dp} is sketched in \textbf{Figure \ref{S2}(a)}.
The optical field $U(\fat{r},0)$ in the plane of interest at $z=0$, is Fourier transformed with the aid of a physical lens, onto the plane of an SLM. The latter modulates this Fourier transform of $U(\fat{r},0)$ with the phase function $\exp(ik_zz)$. This result is then (inverse) Fourier transformed by a second lens to produce the propagated version of the optical field $U(\fat{r},z)$ for a propagation distance of $z$. This propagated output is recorded, using a CCD camera. Since the phase pattern displayed on the SLM can be addressed dynamically, different propagation steps $z$ can be programmed easily and the propagated beam be recorded without moving the camera as with traditional caustic measurements. By way of example \textbf{Figure \ref{S2}(b)} shows the phase patterns displayed on the SLM to achieve beam propagation over distances of $z=0$, $z=\unit[0.5]{m}$, and $z=\unit[1]{m}$. When illuminating such holograms with a fundamental Gaussian beam, one obtains the different propagated versions of the beam that are shown in \textbf{Figure \ref{S2}(c)} at a fixed camera position.

\subsection{Visualizing the concept}

\begin{figure}
\includegraphics{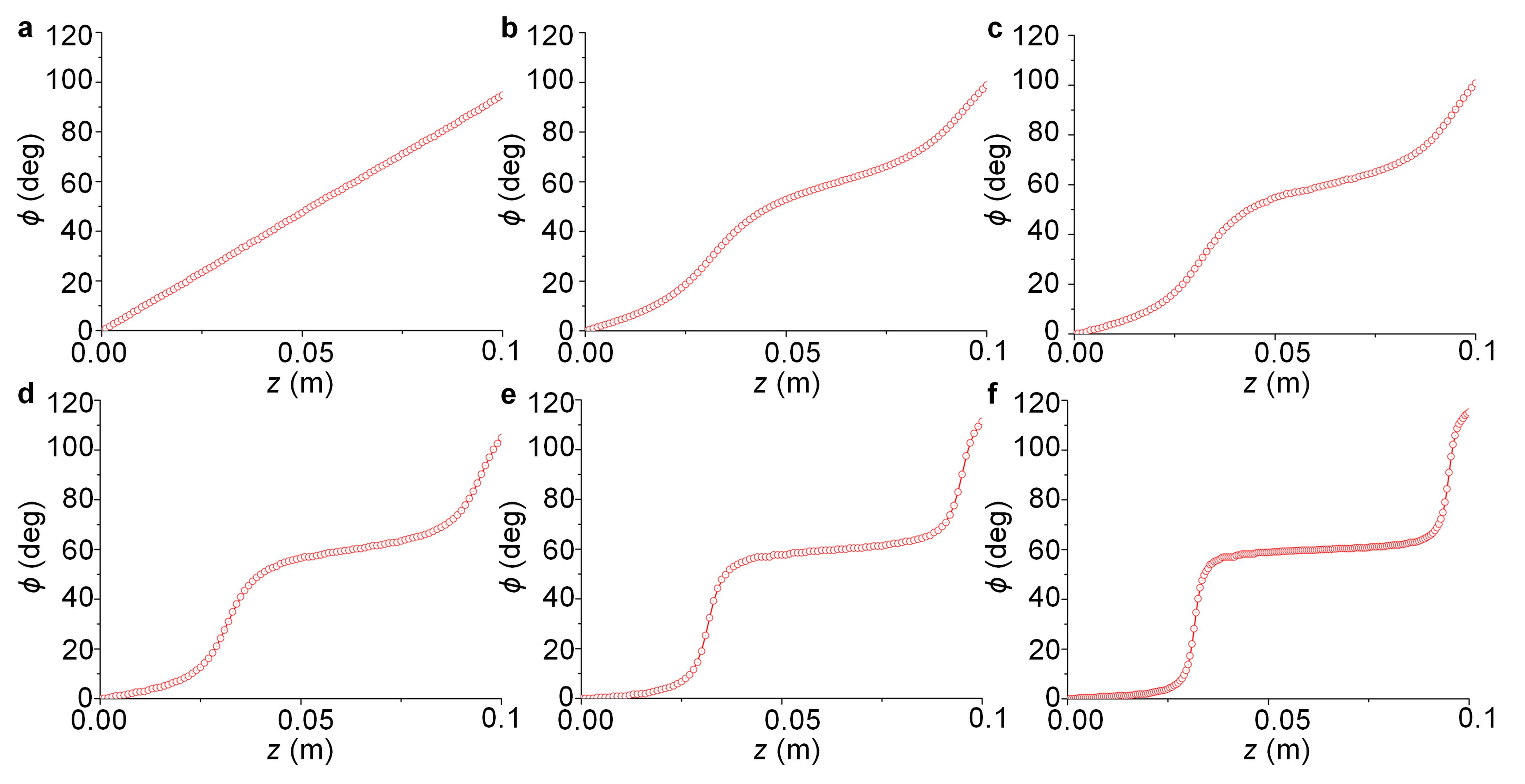}
\caption{Rotation angle $\phi$ as a function of the propagation distance $z$ for different values of the morphology parameter $\theta$. \textbf{a} $\theta=0$, \textbf{b} $\theta=\pi/5$, \textbf{c} $\theta=\pi/4$, \textbf{d} $\theta=\pi/3$, \textbf{e} $\theta=\pi/2.4$, \textbf{f} $\theta=\pi/2.2$. In Media 1-6 the beam intensity is depicted as a function of $z$ for the respective values of $\theta$.}
\label{S3}
\end{figure}

In a superposition of Bessel beams each of the Bessel fields can be given a separate phase factor represented by $k_{z1,2}$. The difference $\Delta = k_{z1}-k_{z2}$ then determines how much the beam rotates during propagation over a distance $z$ and the morphology parameter $\theta$ determines the non-linearity of this rotation. For $\theta=0$, e.g., the rotation is linear. Experimentally, the difference in phase can be tuned by the difference in the radii of the generating ring slits as shown in \textbf{Figure \ref{concept}}. However, in this way the generated Bessel beams would only exist within a defined conical region. To demonstrate the concept of angularly accelerated light, the non-canconical superposition was implemented directly on the SLM, without using ring slits.  Instead we used the complex amplitude modulation technique described by Arrizon, \textit{et al}, \cite{arrizon}. Accordingly, the difference $\Delta$ was set to a certain (nonphysical) value and $z$ tuned while recording the beam with a camera. From each recorded frame the rotation angle of the beam was inferred using the procedure outlined in the Methods. The corresponding results for $l=3$ can be viewed in \textbf{Figure ~\ref{S3}}, depicting the rotation angle as a function of $z$ for different values of the morphology parameter $\theta=\{0,\pi/5,\pi/4,\pi/3,\pi/2.4$,$\pi/2.2\}$, representing rotations of increasing non-linearity. As is seen, for $\theta=0$, the rotation is linear, whereas the angular acceleration at, e.g., $z\approx0.03\,\mathrm{m}$ grows constantly as $\theta$ approaches $\pi/2$, yielding an almost step-like function. Note that the distance within which the beam rotates by a fixed angle does not match the respective distance in \textbf{Figure ~\ref{angrot}}, since $\Delta$ was chosen freely here and has no physical implementation as provided by the ring slits in \textbf{Figure \ref{concept}(c)}. 

\subsection{Rate of rotation}

To determine the rate of rotation, we consider the combination of Bessel beams given in equation (\ref{supu}) at a radius $r=r_0$ where ${\rm J}_{\ell} (r k_{r1})={\rm J}_{\ell} (r k_{r2})$ (there will always be such points), or equivalently a radius that intercepts the peaks of the petals, since ${\rm J}_{\ell} (r k_{r1}) \approx{\rm J}_{\ell} (r k_{r2})$ for small differences in $k_{r n}$. In this case we can neglect the radial dependence and find the intensity to be
\begin{eqnarray}
|g|^2 & = & 2\left[1 + \sin(\theta) \cos(z\Delta) + \cos(\theta) \sin(z\Delta) \sin(2\ell\varphi) \right. \nonumber \\ 
& & \left. + \cos(z\Delta) \cos(2\ell\varphi) + \sin(\theta) \cos(2\ell\varphi) \right] .
\label{inten0}
\end{eqnarray}
Due to the $\varphi$- and $z$-dependence one can see that this intensity pattern rotates. The angular shift, which is obtained by locating the point where the intensity is stationary along $\varphi$ as a function of $z$, is given by equation~(\ref{angle}). The first and second derivatives of this angular shift give the rotation rate [equation~(\ref{rotrate})] and angular acceleration [equation~(\ref{accrate})]. 


To understand the radially dependent rotation rates, we note that the intensity of the field in equation (\ref{inten0}) can be expressed in terms of the sum and difference of the two terms 
\begin{eqnarray}
{\cal I}(r,\varphi,z) & = & g(r,\varphi,z) g^*(r,\varphi,z) 
= {\cal I}_{\rm sum}(r,\varphi,z) + {\cal I}_{\rm dif}(r,\varphi,z) \nonumber \\
& = & \frac{1}{2} R_{\rm s}^2(r) [1+\cos(2\ell\varphi-2\ell\Theta_1)] [1+\sin(\theta)\cos(z\Delta)] \nonumber \\ & & + \frac{1}{2} R_{\rm d}^2(r) [1+\cos(2\ell\varphi-2\ell\Theta_2)] [1-\sin(\theta)\cos(z\Delta)] ,
\label{definte}
\end{eqnarray}
where 
\begin{eqnarray}
R_{\rm s}(r) & = & {\rm J}_{\ell}(r k_{r1}) + {\rm J}_{\ell}(r k_{r2}) \\
R_{\rm d}(r) & = & {\rm J}_{\ell}(r k_{r1}) - {\rm J}_{\ell}(r k_{r2}) \\
\Theta_1 & = & \frac{1}{2\ell} \arctan \left[\frac{\cos(\theta)\sin(z\Delta)}{\sin(\theta)+\cos(z\Delta)}\right] \\
\Theta_2 & = & -\frac{1}{2\ell} \arctan \left[\frac{\cos(\theta)\sin(z\Delta)}{\sin(\theta)-\cos(z\Delta)}\right] \\
\Delta & = & k_{z2}-k_{z1} .
\end{eqnarray}
The sum term represents the part that dominants in the central part of the beam (the inner region, or petals), while the different term covers a larger area of the beam (the outer region, or rings).

From equation (\ref{definte}) we see that the sum and difference terms are each modulated by a $\varphi$-dependent factor and a $z$-dependent factor. The $z$-dependent factors of the two terms are out of phase so that the sum and the difference terms alternate in brightness. The $\varphi$-dependent factors represent a $z$-dependent shift (rotation) in $\varphi$. The rates of the shift (angular velocity) are given by
\begin{eqnarray}
\partial_z \Theta_1 & = & \frac{\Delta}{2\ell}\frac{\cos(\theta)}{1+\sin(\theta)\cos(z\Delta)} \label{velin} \\
\partial_z \Theta_2 & = & \frac{\Delta}{2\ell}\frac{\cos(\theta)}{1-\sin(\theta)\cos(z\Delta)}
\end{eqnarray}
\noindent and similarly the angular accelerations are found to be

\begin{eqnarray}
\partial_{zz} \Theta_1 & = & \frac{\Delta^2}{4\ell}\frac{\sin(2\theta)\sin(z\Delta)}{[1+\sin(\theta)\cos(z\Delta)]^2} \label{accin} \\
\partial_{zz} \Theta_2 & = & \frac{\Delta^2}{4\ell}\frac{\sin(2\theta)\sin(z\Delta)}{[1-\sin(\theta)\cos(z\Delta)]^2}
\end{eqnarray}
Hence, the angular accelerations of the two parts are inversely proportional to their $z$-dependent intensity modulations.  Note that the right-hand sides of equations~(\ref{velin}) and (\ref{accin}) are the same as those of equations~(\ref{rotrate}) and (\ref{accrate}).

In summary, each of the two parts of the intensity pattern given in equation~(\ref{definte}) rotates with a varying rotation rate. When one rotates faster, the other rotates slower. Since the rotation rates are inversely proportional to their respective overall intensities, the one that rotates slow is always brighter than the other one, which rotates fast. 

\subsection{Intensity coupling}
The fact that the two respective terms in equation (\ref{definte}) have varying overall power, implies that they exchange power during propagation. To understand this exchange we use the intensity transport equation \cite{itrans0,itrans,itrans1}
\begin{equation}
\partial_z {\cal I} = \frac{1}{k} \nabla \cdot \left( {\cal I}\nabla\psi \right) ,
\label{itpeq}
\end{equation}
where $\psi$ is the phase of the optical field. The combination ${\cal I}\nabla\psi$ is called the optical current \cite{berryoc} and is proportional to the Poynting vector in the paraxial limit. The intensity transport equation is a statement of energy (power) conservation: the change in the local intensity is balanced by the divergence of the optical current.

Applying the intensity transport equation to the case at hand, we obtain
\begin{eqnarray}
\frac{\Delta}{2}\left[ R_{\rm d}^2(r)-R_{\rm s}^2(r) \right] H(\varphi,z) & = & \frac{\Delta k_{z0}}{2k}\left[ R_{\rm d}^2(r)-R_{\rm s}^2(r) \right] H(\varphi,z)\nonumber \\ & & - \frac{1}{2k} \left[ R_{\rm d}(r) \frac{\partial^2 R_{\rm s}(r)}{\partial r^2} - R_{\rm s}(r) \frac{\partial^2 R_{\rm d}(r)}{\partial r^2} \right] H(\varphi,z) ,
\label{itpeqa}
\end{eqnarray}
where $k_{z0}=k-(k_{z1}+k_{z2})/2$ and
\begin{eqnarray}
H(\varphi,z) & = & \sin(z\Delta)[\sin(\theta)+\cos(2\ell\varphi)]+\cos(z\Delta)\cos(\theta)\sin(2\ell\varphi) \nonumber \\
& = & [1+\sin(\theta)\cos(2\ell\varphi)] \sin(z\Delta+\Theta_3) ,
\label{hdef}
\end{eqnarray}
with 
\begin{equation}
\Theta_3 = \arctan \left[\frac{\cos(\theta)\sin(2\ell\varphi)}{\sin(\theta)+\cos(2\ell\varphi)}\right] .
\label{itpeqb}
\end{equation}

So we see that the intensity transport equation for this case has the form
\begin{equation}
\partial_z {\cal I}_{\rm sum} + \partial_z {\cal I}_{\rm dif} = \alpha \left[ \frac{1}{k} \nabla \cdot ({\cal I}\nabla\psi)_{\rm sum} + \frac{1}{k} \nabla \cdot ({\cal I}\nabla\psi)_{\rm dif} \right] + {\rm coupling~term} ,
\label{itpeqd}
\end{equation}
where $\alpha$ is a factor that reduces the magnitude of the divergence terms, and
\begin{eqnarray}
{\cal I}_{\rm sum} & = & -\frac{\Delta}{2} R_{\rm s}^2(r) H(\varphi,z) \nonumber \\
{\cal I}_{\rm dif} & = & \frac{\Delta}{2} R_{\rm d}^2(r) H(\varphi,z) \nonumber \\
\frac{\alpha}{k} \nabla \cdot ({\cal I}\nabla\psi)_{\rm sum} & = & -\frac{\Delta k_{z0}}{2k} R_{\rm s}^2(r) H(\varphi,z) \nonumber \\
\frac{\alpha}{k} \nabla \cdot ({\cal I}\nabla\psi)_{\rm dif} & = & \frac{\Delta k_{z0}}{2k} R_{\rm d}^2(r) H(\varphi,z) \nonumber \\
{\rm coupling~term} & = & - \frac{1}{2k} \left[ R_{\rm d}(r) \frac{\partial^2 R_{\rm s}(r)}{\partial r^2} - R_{\rm s}(r) \frac{\partial^2 R_{\rm d}(r)}{\partial r^2} \right] H(\varphi,z) .
\end{eqnarray}
Since the coupling term contains both $R_{\rm s}$ and $R_{\rm d}$ in each term, it is responsible for the exchange of intensity between the sum and difference terms during propagation.

Note that the $\varphi$- and $z$-dependent function $H(\varphi,z)$ becomes identical for all terms in the intensity transport equation. It can therefore be cancelled out, leaving only the $r$-dependence
\begin{equation}
R_{\rm d}^2(r)-R_{\rm s}^2(r) = \frac{k_{z0}}{k}\left[ R_{\rm d}^2(r)-R_{\rm s}^2(r) \right] - \frac{1}{k\Delta} \left[ R_{\rm d}(r) \frac{\partial^2 R_{\rm s}(r)}{\partial r^2} - R_{\rm s}(r) \frac{\partial^2 R_{\rm d}(r)}{\partial r^2} \right] .
\label{itpeqe}
\end{equation}
While $H(\varphi,z)$ determines the magnitude and sign of the change in local intensity, the $r$-dependent functions $R_{\rm s}(r)$ and $R_{\rm d}(r)$ govern the coupling between the two regions of the field.

\end{document}